\documentstyle[epsfig,aps,preprint]{revtex}
\tighten
\begin{document}
\draft
\preprint{\vbox{Submitted to Physics Letters B \hfill USC-PHYS-NT-02-99}}

\title{Composite photon and $W^{\pm}, Z^{0}$ vector bosons from a 
%Fermi four-fermion interaction 
top-condensation model 
%of heavy quarks 
at fixed $v = 247$ GeV}
\author{V. Dmitra\v sinovi\' c}
\address{Department of Physics and Astronomy,\\
University of South Carolina, Columbia, SC 29208, USA}
\date{\today}
%\begin{document}
\maketitle
\begin{abstract}
Starting from the historical Fermi four-fermion low-energy effective
electroweak interactions Lagrangian for the third generation of quarks, 
augmented by an NJL type interaction responsible for dynamical symmetry 
breaking and heavy quark mass generation, and fixing the 
scalar (Higgs) field v.e.v. at $v$ = 247 GeV, 
%in the presence of a heavy quark flavour such as the top, 
we show that:
(1) heavy quark bound states $Q \bar Q$ with quantum numbers of the 
$W^{\pm}$ bosons exist 
%in the $J = 1$ scattering amplitude 
for arbitrarily weak (positive) vector coupling $G_F$, so long as the  
quark mass is sufficiently large;
(2) a massive composite neutral vector boson ($Z^0$) appears; 
%with a mass that is
%independent of the weak mixing angle $\theta$.
(3) a massless composite parity-conserving neutral vector boson
%, i.e., the photon 
($\gamma$) appears, the composite Higgs-Kibble ghosts decouple from the quarks 
and other particles,
the longitudinal components of the vector boson propagators vanish, as 
$G_F \to G_{F}^{\rm expt.} = 1/(\sqrt{2} v^2 )$, which also implies that the cutoff 
$\Lambda \to \infty$. Thus, the Fermi interaction model is 
equivalent to a locally gauge invariant theory but with definite values of 
coupling constants and masses. The model discussed here was 
%a simple one
chosen for illustrative purposes
and is {\it not} equivalent to the Standard Model.
\end{abstract}
%\widetext
%\tightenlines

%\section{Introduction}
\paragraph*{Introduction.}

It has long been known that four-fermion contact interactions of the 
Nambu and Jona-Lasinio [NJL] type
can lead to dynamical symmetry breaking along with associated composite
Nambu-Goldstone [NG] bosons \cite{njl61}. Speculations about gauge bosons as
the result of the same or similar mechanism in theories of the Fermi, i.e. 
current $\times$ current type date from the same period \cite{bj63}, but
have generally been regarded as standing on shakier ground. The latter
attempts were concerned with the photon, and not with the $W, Z^0$.
%as modern versions of 
%Jordan's ``neutrino-pair theory of light''. In other words, the 
%photon's masslessness was hoped to be related to the masslessness of the 
%neutrino. 
But the massless photon was attainable only in the $G_V \to \infty$
limit where $G_V$ is the ``vector'' interaction coupling constant of 
dimension mass$^{-2}$, in other words, masslessness had to be introduced by 
hand. It has only recently been shown \cite{vd99} that in models that combine
the NJL type of dynamical symmetry breaking 
and the vector-current Fermi interactions one can have massless
composite (gauge) vector bosons at finite vector coupling $G_V$ as a 
consequence of the renormalization of the Nambu-Goldstone boson decay 
constant $f_{\rm NG}$ which is also the Higgs field vacuum expectation value
[v.e.v.]  $v = f_{\rm NG}$. 
This would be a mere curiosity now if in the meantime the main-stream of 
particle physics had not brought forth precisely such models 
as candidates for the dynamical symmetry breakdown
mechanism in electroweak interactions, going under the generic name of
top-condensation \cite{cvet97}. 

The basic premise of the top-condensation scheme is that the top quark plays a 
special role in electroweak symmetry breaking, due to some, as yet unknown
%unexplained, say ``top-colour'', 
interaction that replaces the elementary Higgs boson sector in the Standard 
Model and is modelled here by an NJL-like quark self-interaction. 
This self-interaction then induces spontaneous symmetry 
breaking with top quark mass generation and the concomitant composite Higgs
boson \cite{cvet97}.
The purpose of this Letter is to demonstrate feasibility of models, not 
merely with a composite Higgs boson, but with composite intermediate vector 
bosons as well, due to such dynamical symmetry breaking. 
We shall do that by way of two illustrative examples - detailed 
predictions of these two models are {\it not} in agreement with
extant data. This is so, because the starting Lagrangians are too
simple. The construction of fully realistic Lagrangians, however, 
is a problem that is shared with other top-condensation models
and is to be addressed in the future. 

As the first example we take the historical Fermi model of low-energy
charged ($V - A$) current weak interactions
%, as established 40 years ago by Feynman and Gell-Mann, and others, 
and show that, when combined 
with an NJL-like model of dynamical symmetry breaking which gives
the heavy quarks their mass, it results in a {\it bound} $Q \bar Q$ vector state
with the quantum numbers of the $W^{\pm}$ bosons for arbitrarily weak
Fermi coupling $G_F \geq 0$, so long as the quark mass $m$ is large enough. 
This is a consequence of keeping $v = f_{\rm NG}$ fixed in the calculation.
We exhibit the dependence of 
the minimal necessary quark mass $m_{\rm min}$ on the Fermi coupling $G_F$. 
We also show that when $G_F$ exceeds $0.75/(\sqrt{2} v^2 )$ the charged vector 
bound state exists for all values of the quark mass $m$. 
This bound state's mass approaches 
%the limit of 
$\sqrt{3} m$ as $G_F \to G_{F}^{\rm expt.} = 1/(\sqrt{2} v^2 )$, which also
corresponds to the cutoff $\Lambda \to \infty$.
In this ``continuum'' limit we show that the Higgs-Kibble ghosts 
(the ``would-be NG bosons'') decouple
from the rest of the theory, the longitudinal components of the vector boson 
propagators vanish
and the theory attains the form of a local gauge theory in the unitary gauge, but 
with definite predictions for the values of Higgs- and vector boson masses. 
%We interpret the former as
%a massless vector gauge boson (``$\gamma$'') and the latter as 
As the second example we take the weak neutral current (WNC) 
low-energy effective Fermi Lagrangian. A completely analogous analysis then leads 
%Thus we have been led 
to the following conclusions:
1) A massive parity-violating neutral vector state (``$Z^0$'') whose mass 
$M_{Z} \to \sqrt{2} M_{W}$ is 
independent of the mixing angle $\theta$, however. 
2) A light parity-conserving neutral vector state (``$\gamma$'') whose mass 
approaches zero as $G_F \to G_{F}^{\rm expt.} = 1/(\sqrt{2} v^2 )$. Many of 
the phenomenological deficiencies of these models can be overcome by using more
complicated Lagrangians, see Ref. \cite{cvet97}, which we shall avoid here so 
as not to obscure the
more important conceptual developments, {\it viz.} the very existence of bound
state intermediate vector bosons, and their properties in the continuum limit.

\paragraph*{Charged currents}
For concreteness' sake and as an historical curiosity \footnote{It seems
%worth pointing out 
interesting that even the Fermi four-point interaction can lead to (composite)
$W^{\pm}$ intermediate vector bosons, without ever introducing Yang-Mills fields.} 
we take the charged-current [CC] weak interactions effective Fermi
field theory augmented by the NJL Lagrangian 
%${\cal L}_{\rm CC}$ 
\begin{eqnarray} 
{\cal L}_{\rm CC} &=& 
\bar{\psi} \big[{\rm i} {\partial{\mkern -10.mu}{/}} ]\psi
+ G_{1} \Big[ (\bar{\psi} \psi)^2 + 
  (\bar{\psi} {\rm i} \gamma_5 \mbox{\boldmath$\tau$} \psi)^2 \Big] 
\nonumber \\ 
&-& {G_{F} \over{4 \sqrt{2}}} 
\Big[ (\bar{\psi} \gamma_{\mu} \left(1 - \gamma_5 \right) 
\mbox{\boldmath$\tau$} \psi)^2 - 
(\bar{\psi} \gamma_{\mu} \left(1 - \gamma_5 \right) 
\mbox{\boldmath$\tau$}^{3} \psi)^2 \Big]\: ,
\label{e:lag2}  
\end{eqnarray} 
as our starting point, 
where $\psi$ is a Dirac field doublet of the top and bottom quarks.
\footnote{The question arises if this model works with the first, or the 
second generation quarks? The answer is in the negative, 
%as a matter of principle, 
because: (i) the associated would-be NG bosons [$\pi, K, D, \eta, \eta^{'}, 
\eta_{c}$] have all been observed as stable particles, and 
(ii) all of these ``would-be NG bosons'' are massive; both points 
being in conflict with the notion that these states should be {\it massless}
Higgs-Kibble ghosts that {\it must disappear in the unitary gauge}, see text 
below. The $t \bar Q$ ground states [$J^P = 0^-$], where $Q = t, b$, have 
not been found, however, 
in searches at Fermilab; this fact can be interpreted as being in 
accord with the top-condensation hypothesis, which assigns the
role of Higgs-Kibble ghosts to these states.}
This model exhibits dynamical symmetry breakdown into a nontrivial ground 
state with quark mass generation due to the NJL self-interaction term ($\sim G_1$), 
when treated non-perturbatively. This self-interaction is postulated 
as a model of some, as yet unknown high-energy interactions 
%(hypercolour, top-colour)
responsible for the electroweak symmetry breaking.
For the sake of clarity we chose the simplest possible NJL self-interaction.
As a consequence of this choice the top and bottom quarks have equal masses, 
$m_t = m_b$, in stark contrast with experiment
[$m_{t} \simeq$ 175 GeV, $m_{b} \simeq $ 5 GeV].
The realistic mass splitting can be reproduced in our model, but at the price of 
a more complicated NJL-like Lagrangian \cite{cvet97}, which we avoid using here
for clarity's sake. 
\footnote{
In a more realistic model there would also be gluon fields coupled to the
quarks, but we neglect them here, as the mean loop momenta are comparable 
to the top quark mass and asymptotic freedom of QCD reduces their contribution.}
For this reason we do not expect phenomenological success of this model, and 
shall be content to demonstrate the existence of {\it bound state} intermediate 
vector bosons, our main result here. 

The non-perturbative dynamics of the model, to
leading order in $1/N_c$, are described by two Schwinger-Dyson [SD]
equations: the gap equation and the Bethe-Salpeter [BS] equation. Our model
has three parameters: two positive coupling constants $G_1 , G_F$ of dimension 
(mass)$^{-2}$ and a regulating cutoff  $\Lambda$ that determines the mass scale.
% are the model parameters. $G_1$ and the cutoff $\Lambda$  
The gap equation establishes a relation
between  the constituent quark mass $m$ and the two free parameters $G_1$ and 
$\Lambda$. 
This relation is not one-to-one, however: there is a (double) continuum of allowed
$G_1$ and $\Lambda$ values that yield the same nontrivial solution $m$ to the 
gap equation. As discussed in Ref. \cite{vd99}, 
one of the two continuum degeneracies can be eliminated by fixing the value of 
the Nambu-Goldstone [NG] boson
decay constant $f_{\rm NG}$ at the observed value 247 GeV. 
There is a renormalization of the ``bare'' ($G_F = 0$) pion decay 
constant $f_{0}$ to $v$ by a factor of $\sqrt{Z}$ according to
\begin{equation}
Z^{-1} = 1 - \sqrt{2} G_F v^{2} = \left({v \over f_{0}} \right)^{2} .
\label{e:ga1} 
\end{equation}
%between $Z$ and $G_F$ and $f_{\rm NG}$, the last of which
where $v$ is kept constant.
An important consequence of the relation (\ref{e:ga1}) 
%and of the second line of Eq. (\ref{e:ga}) 
is the inequality $0 \leq Z^{-1} \leq 1$.
This is equivalent to an upper bound on $G_F$:
\begin{equation}
G_F  \leq 1/(\sqrt{2} v^{2})  ~,
\label{e:gaineq} 
\end{equation}
apart from the trivial lower bound $G_F \geq 0$. 
Now, from the Goldberger-Treiman
(GT) relation $f_{0} g_{p} = m$ one finds 
\begin{equation}
\left({v \over m} \right)^{2} = 
{3 Z \over{(2 \pi)^{2}}} \sum_{s = 0}^{2} C_{s} \log(M^2_s/m^2)~ ,
\label{e:gt} 
\end{equation}
where the $C_s$ and $M^2_s = m^2 + \alpha_s \Lambda^2$
are the standard parameters of the Pauli-Villars (PV) regularization
scheme \cite{iz80}. 
The result of the constraint Eq. (\ref{e:gt}) is a family of quark mass 
$m$ vs. cutoff $\Lambda$ curves, one for every value of $Z$, or $G_F$, 
shown in Fig. 1, all points on which satisfy $v$ = 247 GeV. 
Note that $\Lambda \to \infty$
as $Z \to \infty$, for any given value of $m$. In other words the 
``critical point'' limit $Z \to \infty$ is equivalent to the 
``continuum limit'' $\Lambda \to \infty$ at fixed $v$ and $m$.

\paragraph*{Solutions to the BS equation}
The BS equation in the charged-current channel reads
\begin{eqnarray}
1 + {G_F \over{2\sqrt{2}}} \left[\Pi_{V}(s_{W}) + \Pi_{A}(s_{W})\right] = 0
\label{e:bse}
\end{eqnarray}
In order to find the bound-state roots $0 \leq s_{W} \leq 4 m^2$ to this 
equation we use the polarization functions $\Pi_{V,A}$ shown in Ref. \cite{vd99}. 
If we keep $m$ and $v$ (hence also $g_{\rm PS}$) fixed, then $\Pi_{V,A}(s), F(s)$ 
are functions of $G_F$:
\begin{eqnarray}
\Pi_V(s,G_F) &=&
- {2 Z \over{3}} g_{\rm PS}^{-2} 
\left[ s + \left(s + 2 m^2\right)[F(s) - 1] \right] 
\nonumber
\\
\Pi_{A}(s,G_F) &=& \Pi_V(s)+ 4 Z v^2 F(s)
\nonumber
\\
F(s,G_F) &=& 1 - \frac{3 g_{\rm PS}^2}{2 Z \pi^2} \{\sqrt{- f}
{\rm Arccot}{\sqrt{-f}} - 1 \}_{PV}
~,
\label{e:pi1}
\end{eqnarray}
where
$f = 1 - 4m^2/s$, Pauli-Villars (PV) regularization of $F(s)$ has been used,
and we kept $Z^{-1}$ as an abbreviation for $1 - \sqrt{2} G_F v^{2}$,
for the sake of conciseness.
The analytic approximation to the $W$ boson mass reads
\begin{equation}
M_{W}^{2} = {3 g_{\pi}^{2} \over{Z \sqrt{2} G_{F}}} + 3 m^2
%\left(^{-1} + 1 \right)
= 3 m^{2} \left({Z^{-1} \over{1 - Z^{-1}}} + 1\right) 
= {3 m^{2} \over{1 - Z^{-1}}}  ~.
\label{e:mv} 
\end{equation}
According to Eq. (\ref{e:mv}) the bound-state ought to dissolve for 
$Z \leq 4$, but the numerical solution of the BS Eq. (\ref{e:bse}) 
shows that the bound-state may exist 
at even lower values of $Z$, i.e., at lower values of $G_F$, depending on
the value of the (heavy) quark mass $m$.
In the next section we shall find the range of values of $m = m(\Lambda)$ 
in which there is a bound state for a given $G_F$. 

\paragraph*{Minimal quark mass necessary for a $W^{\pm}$ bound state}
In order to determine the values of $G_F, m$ for which the charged current
BS Eq. (\ref{e:bse}) has bound-state solutions it is sufficient to consider the
inequality
\begin{eqnarray}
1 + {G_F \over{2 \sqrt{2}}} 
\left(\Pi_{V}(4 m^{2}) + \Pi_{A}(4 m^{2})\right) \leq 0~.
\label{e:ineq}
\end{eqnarray}
Use Eq. (\ref{e:pi1}) to find
\begin{eqnarray}
\Pi_V(4 m^2) + \Pi_A(4 m^2) &=&
{4 \over{3}} Z v^{2} \left[2 - 3 F(4 m^2) \right] 
\nonumber \\
&=& 
- {4 \over{3}} Z v^{2}
\left[1 + 2 Z^{-1} \left(\frac{3 g_{\rm PS}}{2 \pi}\right)^2 \right] ~.
\label{e:pi2}
\end{eqnarray}
This and the inequality (\ref{e:ineq}) lead to
\begin{eqnarray}
\left(\frac{3 g_{\pi}}{2 \pi}\right)^2 &\geq&
{4 - Z \over{2 (1 - Z^{-1})}} ~,
\label{e:ineq2}
\end{eqnarray}
which is our vector bound-state criterium. For $Z \geq 4$ the r.h.s. 
of this inequality is non-positive, i.e., 
the inequality is trivially satisfied and there is a vector bound 
state for all real values of $m$. For $Z \leq 4$ this turns into a
lower bound on the quark mass $m$:
\begin{eqnarray}
m &\geq& m_{\rm min}(Z) = v \frac{2 \pi}{3} 
\sqrt{{4 - Z \over{2 (1 - Z^{-1})}}} ~.
\label{e:ineq3}
\end{eqnarray}

\paragraph*{Decoupling of the longitudinal components in the vector propagators 
and of Higgs-Kibble ghosts at the critical point}
The total NG boson $Q{\bar Q}$ coupling constant $g_{\rm NG}$ is the sum of 
pseudoscalar (PS) and pseudovector (PV) coupling that can be written as a 
function of the renormalization constant $Z$:  
\begin{equation}
g_{\rm NG} = Z^{-1/2} g_{\rm PS} = Z^{-1/2} \left({m \over v} \right)~.
\label{e:gh} 
\end{equation}
where $g_{\rm PS} = m/ v$ is the pseudoscalar coupling. Hence it follows that 
NG boson $Q{\bar Q}$ coupling vanishes in the critical limit $Z \to \infty$. 
This shows that 
the NG bosons, a.k.a. Higgs-Kibble ghosts decouple at the critical point.

The longitudinal part of the axial vector propagator is proportional to
\begin{eqnarray}
D_{W}^L(s)
&=&
{1 \over 2} \left[\left(A + C\right) - \sqrt{\left(A - C\right)^2 + 4 B^2}\right]
\nonumber \\
A \pm C &=& 
Z^{-1} g_{\rm PS}^{2} \left[1 + \sqrt{2} G_F f_{0}^2 F(s) \right] 
\pm {s \over{2 \sqrt{2}}} G_F F(s)
\nonumber \\
B &=& 
%Z^{-1} g_{\rm PS}^{2} \left[1 + \sqrt{2} G_F F(s) \right] \pm 
{1 \over{\sqrt{2}}} G_F m \sqrt{s} F(s)~.
\label{e:abc}
\end{eqnarray}
Taking the critical point limit we find
\begin{eqnarray}
A \pm C &=& 
g_{\rm PS}^{2} F(s) \left[1 \pm {s \over{4 m^{2}}}  \right] 
\nonumber \\
B &=& 
g_{\rm PS}^{2} F(s) {\sqrt{s} \over{2 m}} ~,
\label{e:abc1}
\end{eqnarray}
and consequently
\begin{eqnarray}
D_{W}^L(s) &=& 0~.
\label{e:abc3}
\end{eqnarray}
Thus we see that the longitudinal parts of the $W$ propagators vanish at the 
critical point. In other words, Fermi theory results at the critical point
$\Lambda, Z \to \infty$ are much like those of a spontaneously broken local 
gauge field theory
with a massive $W$ [of mass $\sqrt{3} m$] due to the Higgs mechanism, and
a Higgs mass of $2m$.  Note that both $M_{W}$ and $M_{\rm Higgs}$ are finite 
in the $\Lambda \to \infty$ limit.

\paragraph*{Neutral currents and the photon}
Having shown that the Fermi charged current interactions lead to the existence of
$W^{\pm}$ bosons provided that the symmetry is dynamically broken, we ask ourselves
if the same is true for the weak neutral current [WNC].
We start from the empirical WNC effective low-energy Fermi interactions
\begin{eqnarray} 
{\cal L}_{\rm WNC} &=& 
%\nonumber \\ &-& 
- {G_{F} \over{\sqrt{2}}} \Big[ 
\bar{\psi} \gamma_{\mu} \left(1 - \gamma_5 \right) 
{\mbox{\boldmath$\tau$}^{3} \over 2} \psi 
- 2 \sin^{2} \theta J_{\mu}^{\rm EM} \Big]^2 \:; 
\label{e:lag1}  
\end{eqnarray}  
with the same kinetic term and ``Higgs sector'' as for the charged currents, Eq. 
(\ref{e:lag2}). We use the same non-perturbative approximation described by 
the gap equation and the BS equation (\ref{e:bse}). 
The latter is a matrix equation in flavour space now, 
\begin{eqnarray}
\left(1 + {G_F \over{2\sqrt{2}}} \mbox{\boldmath$\Pi$}_{WNC}(s)\right)^{-1} T(s)
 = 0~,
\label{e:bse1}
\end{eqnarray}
due to the flavour-mixing terms in the WNC Lagrangian (\ref{e:lag1}). Here
\begin{eqnarray}
{\bf 1} + {G_F \over{2\sqrt{2}}} \mbox{\boldmath$\Pi$}_{WNC} =
\left(\begin{array}{cc}
{\left[1 + {G_F \over{2\sqrt{2}}} \left(\mbox{\boldmath$\Pi$}_{A} 
+ \cos^2 2\theta \mbox{\boldmath$\Pi$}_{V}\right)\right]} & 
{- \alpha {G_F \over{2\sqrt{2}}} \mbox{\boldmath$\Pi$}_{V}} \\
{- \alpha {G_F \over{2\sqrt{2}}} \mbox{\boldmath$\Pi$}_{V}} &
{\left[1 + {G_F \over{2\sqrt{2}}} \left({2 \over 3} \sin^2 \theta\right)^2
 \mbox{\boldmath$\Pi$}_{V}\right]}~,
\end{array}\right)
\end{eqnarray}
and
\begin{eqnarray}
\alpha = {2 \over 3} \sin^2 \theta \cos 2\theta~.
\end{eqnarray}
The masses of the states are determined by the roots of the secular equation:
\begin{eqnarray}
{\cal{D}}(s)=
Det\left(1 + {G_F \over{2\sqrt{2}}} \mbox{\boldmath$\Pi$}_{WNC}(s) \right) = 0
\end{eqnarray}
We evaluate the l.h.s. at the critical point and find the quadratic polynomial
\begin{eqnarray}
{\cal{D}}(s) \simeq s
\left( s - 6 m^2 \right) = 0~,
\end{eqnarray}
that has one massless and one massive root. 
Thus we have been led to the conclusion that there is:
1) A massive parity-violating neutral vector state (``$Z^0$'') whose mass is 
larger by factor of $\sqrt{2}$ than that of the $W$ and is independent of the 
mixing angle $\theta$. 
2) A light parity-conserving neutral vector state (``$\gamma$'') whose mass 
approaches zero as $G_F \to 1/(\sqrt{2} v^2 )$.
We see that the resulting ``$\gamma - Z^0$'' sector of the Fermi theory is 
{\it not} identical to the $\gamma - Z^0$ 
sector of the Standard Model, due to 
the fact that we started from the pure WNC effective Fermi interaction, i.e.,
without taking into account the EM interaction.
% which ought to be another result of mixing. 
This we did for the sake of simplicity and concreteness.
The point of this Letter is the demonstration of feasibility of models with
composite Higgs and intermediate vector bosons due to dynamical symmetry breaking, 
not the construction of the most realistic such model.
%The latter is a problem for the future.
This paper also seems to be the first demonstration
%, and in a sense a fullfilment 
of the Bjorken--Bialynicki-Birula programme \cite{bj63} at finite $G_V$. 
No massless fermions are necessary, but a dynamical symmetry breaking and 
the unification of EM and weak interactions seem indispensable.

\paragraph*{Conclusions}
We have shown that the Fermi model of electroweak interactions augmented 
with an NJL self-interaction of third generatio quarks
%responsible for the condensation of top 
binds $Q \bar Q$ (where $Q = t, b$) 
pairs into $W^{\pm}$ vector boson states in the charged current sector
for arbitrarily small positive values of the coupling constant $G_F$, 
provided the top/bottom quark mass $m$ is large enough, and for all values 
of $m$ with $G_F \geq 0.75 G_{F}^{\rm expt.} = 0.75 (\sqrt{2} v^2)^{-1}$, 
as a consequence of keeping $v$ fixed at 247 GeV. 
This implies that bound-state $W^{\pm}$ vector bosons can be found at as low 
values of $G_F$ as 0.75 of the observed one. The neutral current boson 
remains unbound 
%for all values of $G_F$ 
in the simplest WNC model, but the 
photon is a bound state that becomes massless as
$G_F \to G_{F}^{\rm expt.} = (\sqrt{2} v^2)^{-1}$, i.e., as
$\Lambda, Z \to \infty$.
An analog of exact local gauge symmetry is recovered at this critical point. 
This seems to be 
the first demonstration of photon's masslessness and of bound-state nature of 
the $W$ in this sort of model. The lack of conformity between
%the $\gamma - Z$ sectors of 
this and the Standard Model is a matter of
not having started from an optimal four-fermion interaction.
% term corresponding to the EM interaction. 
Work on a realistic Lagrangian that would remedy this and other
phenomenological deficiencies is in progress. 

\paragraph*{Acknowledgements}
The author would like to thank K. Kubodera and F. Myhrer for discussions.
%and comments on the manuscript.

\begin{figure}
\begin{center}
\epsfig{file=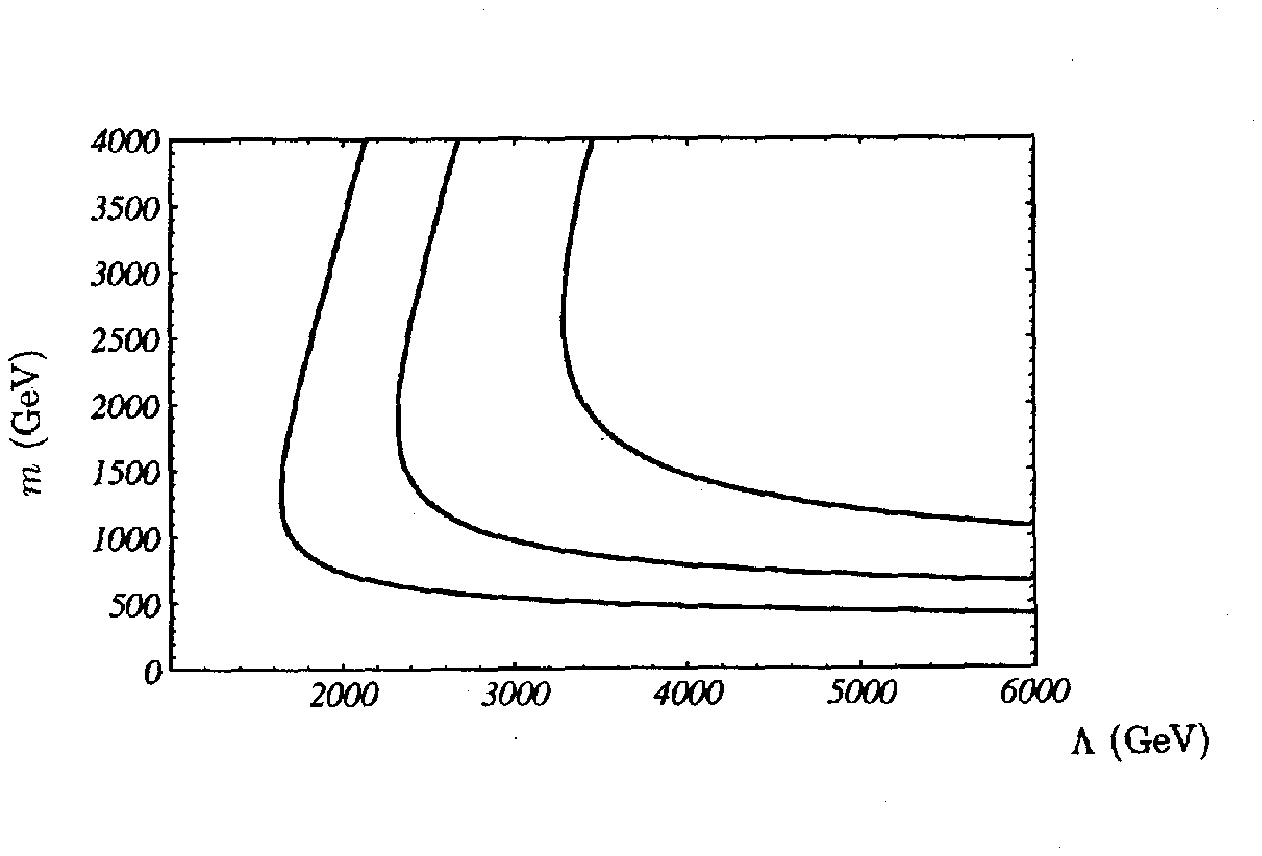,width=18cm} 
\end{center} 
\caption{
The top, or bottom quark mass $m$ (in units of GeV) as a function of the 
Pauli-Villars 
cutoff $\Lambda$ (in units of GeV) at fixed $v$ = 247 GeV in our model: 
$Z = 1, 2, 4$ - the far left h.s.-, the middle- and the far right h.s. curves, 
respectively.}
\label{f:1}
\end{figure}
\end{document}